\documentstyle[epsfig]{aipproc}
\begin{document}
\title{Investigations of  Positron Annihilation Radiation} 
\author{P.A.~Milne$^{1}$, J.D.~Kurfess$^{2}$, R.L.~Kinzer$^{2}$, 
M.D.~Leising$^{3}$ and D.D.~Dixon$^{4}$}

\address{
$^{1}$NRC/NRL Resident Research Associate,
Naval Research Lab, 
\linebreak
Code 7650,
Washington DC 20375
\linebreak
$^{2}$Naval Research Lab, Code 7650,
Washington DC 20375
\linebreak
$^{3}$Clemson University, Clemson, SC 29631
\linebreak
$^{4}$Institute of Geophysics and Planetary Physics,
 Univ. of Cal. Riverside, CA 92521
}

\maketitle

\begin{abstract}
By combining OSSE, SMM and TGRS observations of the galactic 
center region, Purcell et al. (1997) and Cheng et al. (1997) 
produced the first 
maps of galactic positron annihilation. That data-set has been augmented 
with additional data, both recent and archival, and re-analyzed
to improve the spectral fitting.\footnote{This paper represents the 
combination of presentations D-5 and D-12.} The improved spectral fitting has
enabled the first maps 
of positronium continuum emission and the most extensive maps of 511 keV
line emission. Bulge and disk combinations have been compared with 
the 511 keV line data, demonstrating that extended bulges 
are favored over a GC point source
for every disk model tested. This result is independent of whether
OSSE-only, OSSE/SMM, or OSSE/SMM/TGRS data-sets are used.
The estimated bulge to disk ratio (and to a lesser extent the total
flux) is shown to be dependent upon the assumption of bulge shape.
A positive latitude enhancement is shown to have an effect upon
the B/D ratio, but this effect is secondary to the choice of bulge shape. 
\end{abstract}

\section{Introduction}

One of the primary objectives of the OSSE instrument on 
NASA's COMPTON Observatory has been to 
understand the nature of galactic positron 
annihilation radiation. 
Through 8$\frac{1}{2}$ years of observations, 511 keV line 
emission has been detected, but the emission has never been  
unambiguously attributable to a given discrete source.
The galactic center (GC) region's 511 keV emission was monitored
by the Gamma-Ray Spectrometer on-board the Solar
Maximum Mission (SMM) (1980-1988), the Transient Gamma-Ray
Spectrometer (TGRS) on-board the WIND mission (1995-1997) and
with multiple OSSE observations (1991-present). None of these detections
require variable sources in addition to the two component 
models discussed here to explain the measured fluxes (Share et al. 1990,
Harris et al. 1998, Purcell et al. 1997).  These results have supported
 the suggestion that the majority of the emission is
diffuse.
Reported here are preliminary results 
of the extension of the OSSE analysis into three new 
areas: the inclusion of observations in regions with no 
$a \hspace{1mm} priori$ expectations of positron annihilation radiation, 
the extension of the analyzed region to include a larger fraction of 
the Galaxy, and the mapping of the positronium continuum 
component (PCONT) 
of the total annihilation emission. 
The derived PCONT flux values have a stronger dependence upon fitting 
the underlying continuum 
than do the 511 keV line flux values, and thus 
detailed analyses (model-fitting \& 1D cuts) are only performed upon 
the 511 keV data-set in this preliminary presentation.\footnote{Among the 
potentially important effects not yet addressed are 
emission from the diffuse cosmic-ray continuum and a correction for  
scan-angle dependent background.}

Previously published OSSE results have focussed upon the 511 keV 
line emission emanating from the central radian of the inner Galaxy, 
mapping emission from $|l|$ $\leq$ 33$^{\circ}$, 
$|b|$ $\leq$ 17$^{\circ}$, and model-fitting on a 
$|l|$ $\leq$ 90$^{\circ}$,$|b|$ $\leq$ 
45$^{\circ}$, 1$^{\circ}$x1$^{\circ}$
grid. In a result first reported
 at the Fourth Compton Symposium and based on 6+ years of OSSE data, 
Purcell et al. 1997 (hereafter PURC97), showed evidence for three components 
to the 511 keV line emission, (1) an intense slightly 
extended emission centered in the 
direction of the GC, (2) a fainter planar emission, 
and (3) an 
unexpected enhancement of emission from positive latitudes (PLE). 
To generate that data-set, two basic types of data 
were used; (1) standard \& offset pointing
 data where the scan angle crossed the 
GC and/or was perpendicular to the galactic plane,
 and (2) mapping data which searches 
for emission by observing a large sky region at regularly spaced 
intervals along a scan path.\footnote{See 
PURC97 for details of the OSSE pointing strategies 
and background techniques.} The live-time from mapping observations is 
spread over many more pointings (16 or 32 for mapping versus 3-9 
for standard), so the sensitivity 
per source pointing achieved is inferior to
 the standard or offset observations. 

The present work relaxes all selection criteria, initially including all 
observations whose source and background pointings are within the  
$|l|$ $\leq$ 90$^{\circ}$, $|b|$ $\leq$  
45$^{\circ}$ region. To extract the 
positronium component from the total emission, three spectral models 
have been fit to each spectrum (from 60 keV to 700 keV); (1) a single power-law 
+511 keV line + PCONT (as fit by PURC97, though they fit from 50 keV -4 MeV), 
(2) a power-law with an exponential fall-off +511 keV +PCONT, 
(3) a thermal bremsstrahlung model 
+511 keV +PCONT. If the best-fitting model (of the 
three) is deemed to provide an acceptable fit, then that result is 
selected for subsequent mapping analysis. This has been
 done for the 1153 observations in the data-set. Fewer 
than 20 observations have been rejected in this preliminary, all-inclusive 
analysis. The combination of including archival data and the use of new 
data collected in the 2+ years since the PURC97 paper has increased 
the  total GC exposure from 1.9 x 10$^{7}$ det$\cdot$s to 
8.6 x 10$^{7}$ det$\cdot$s, as seen in Figure 1. As a result of this 
exposure, the fraction of the GC 
region ($|l|$ $\leq$ 90$^{\circ}$, $|b|$ $\leq$
45$^{\circ}$) mapped above our exposure 
threshold has increased from 22\% to 83\%. 

\begin{figure}
\label{exp}
\centerline{\epsfig{file=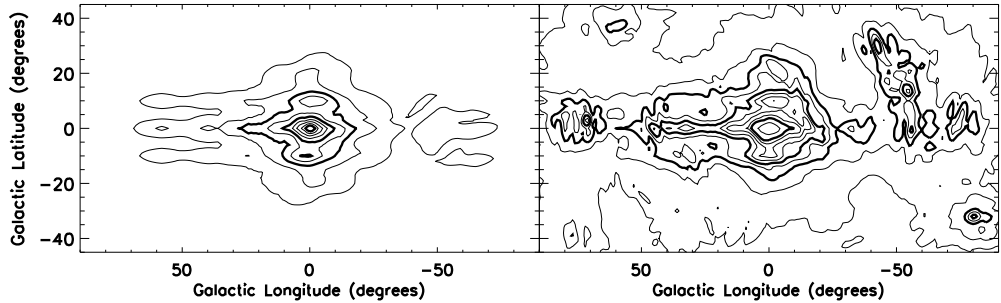, width=6.0in}}
\vspace{2mm}
\caption{The exposure map of the expanded data-set compared
with the earlier data-set of Purcell et al. (1997). The maps are in units of
10$^{9}$ cm$^{2}$ s, with the contour levels 0.05, 0.25, 0.5, 0.75, 1, 1.25,
1.5, 1.75, 2, 2.5.
}
\end{figure}

The combined OSSE/SMM data-sets are used rather than the
OSSE-only data-set because the OSSE background-subtraction
 technique leads to
differential rather than absolute fluxes. OSSE is insensitive to
both isotropic emission and modest intensity gradients. The SMM
fluxes are not absolute either, being insensitive to isotropic emission. 
Assuming isotropic emission to be zero, the SMM data contributes 
an overall normalization (Share et al. 1988). The
SMM FoV is wide ($\sim$130$^{\circ}$ FWHM), but it does contribute 
limited spatial
information as the response peak swept through
the GC region along the ecliptic. One important difference between
the PURC97 data-set and this one is the use of TGRS data. PURC97 used
the TGRS data from Teegarden et al. (1996). That data has since been
re-analyzed by Harris et al. (1998), with the resultant 511 keV line
flux reduced by almost 20\%. The Harris et al. (1998) data-set is 
used for the model-fitting studies shown in Table 1. 
SVD maps of the combined OSSE/SMM/TGRS
data-set were not ready for these proceedings, but a RL map of the 
OSSE/SMM/TGRS data-sets (not shown) is in general 
agreement with the OSSE/SMM map. 

\section{Mapping Positron Annihilation Emission}

Two techniques have been employed to map the OSSE data-set:  
minimizing $\chi^{2}$ with an adaptation
of the Richardson-Lucy Algorithm (RL), and
response matrix
inversion using truncated Singular Valued Decomposition (SVD). SVD was
described in PURC97, RL will be described in detail in an upcoming
work.\footnote{In
short, RL adds flux distributed according to 
the instrument response to the source/background regions
to raise/lower individual flux values to match each observation. Through
successive iterations, map structure develops and the overall $\chi^{2}$
lowers.} Differences between the two resulting maps suggest the level 
of the uncertainties involved. 

\begin{figure}
\label{maps}
\centerline{\epsfig{file=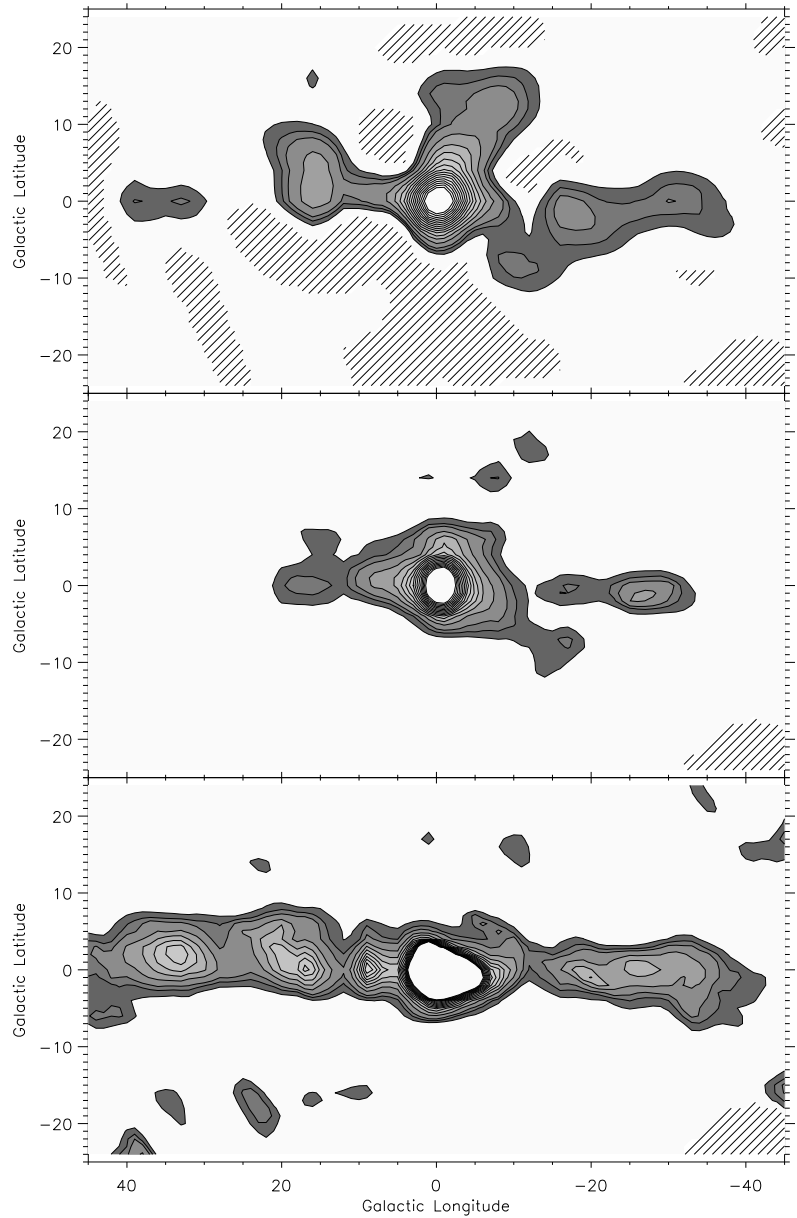, width=11cm}}
\vspace{-.70cm}
\caption{The SVD-511 map of the OSSE/SMM data-set 
is shown in figure 2a, the RL-511
map is shown in figure 2b. The RL-positronium continuum
 map of the OSSE-only data-set 
is shown in figure 2c. All
contours are in increments of 2.3 x 10$^{-3}$ phot cm$^{-2}$ s$^{-1}$
ster$^{-1}$. For the RL maps, the hatched region is where the
OSSE exposure does not meet a
minimum threshold, for the SVD map the regions where
unphysical negative fluxes occur is also hatched.}
\end{figure}
                     
The upper and middle panels of 
Figure 2 shows the SVD and RL 511 keV maps of the 
combined OSSE and SMM data-sets. Both maps show 
three principle features, intense emission centered near the GC 
(hereafter called bulge emission), a fainter planar emission 
(hereafter called disk emission), and emission from the negative 
longitude/positive latitude region (hereafter called a PLE). 
The PLE is more pronounced in the SVD map than in the RL map. 
The hatched region does not meet a minimum exposure threshold 
(or maps negative intensity). We emphasize the level 
to which the emission is concentrated in these three components. Although a 
far larger fraction of the inner radian is mapped than in PURC97, there 
is no indication of intense emission from any of these newly mapped 
regions. The apparent dominance of the bulge and disk emissions along with a 
contribution from the PLE  
drive the use of two and three component models when model-fitting.

Shown in the lower panel of Figure 2 
 is a RL map of the fitted positronium continuum 
fluxes of the OSSE data-set. In 
most astrophysical environments, more PCONT photons are produced in 
annihilation events than 511 keV line photons. For a positronium 
fraction of 0.95, the PCONT:511 ratio is 3.7:1.\footnote{OSSE has 
measured f$_{Ps}$=0.97$\pm$0.03 at the GC (Kinzer et al. 1996).  
This value is consistent with the TGRS measured value, 
f$_{Ps}$=0.94$\pm$0.04 (Harris et al. 1998).} As a result, the PCONT 
map is more intense. As explained in the introduction, 
this map is preliminary as a number of potential 
biases have not yet been addressed. Nonetheless, the 
dominance of intense bulge and fainter 
disk components appears to agree with the 511 keV maps.
The principal difference is the lack of evidence of PLE emission, 
as will be discussed in the next section.

\section{Longitude and PLE Cuts}

A measure of the planar structure in the 511 keV emission 
can be seen by taking a cut along the galactic plane of the 
RL and SVD maps. Cuts for the RL and SVD maps, as well as a 
model combining an R$^{1/4}$ bulge and the DIRBE 100 disk are 
 shown in the upper panel of Figure 3.
 Each is the sum of the $|b|$ $\leq$ 2$^{\circ}$ pixels. 
The SVD data is plotted with 
1$\sigma$ error bars per degree. The SVD and RL 511 maps show rough 
agreement, except in the  +18$^{\circ}$ to +27$^{\circ}$ region, where 
the peaks and valleys are exaggerated in the SVD map relative to the 
RL map. The model is more centrally peaked and the centroids of the 
maps are slightly offset towards negative longitude, but general 
agreement exists between the model and the maps. 
A systematic survey of the inner galactic plane scheduled for CGRO 
Cycle 9 will improve the sensitivity of the longitudinal cut. 

\begin{figure}
\label{cuts}
\centerline{\epsfig{file=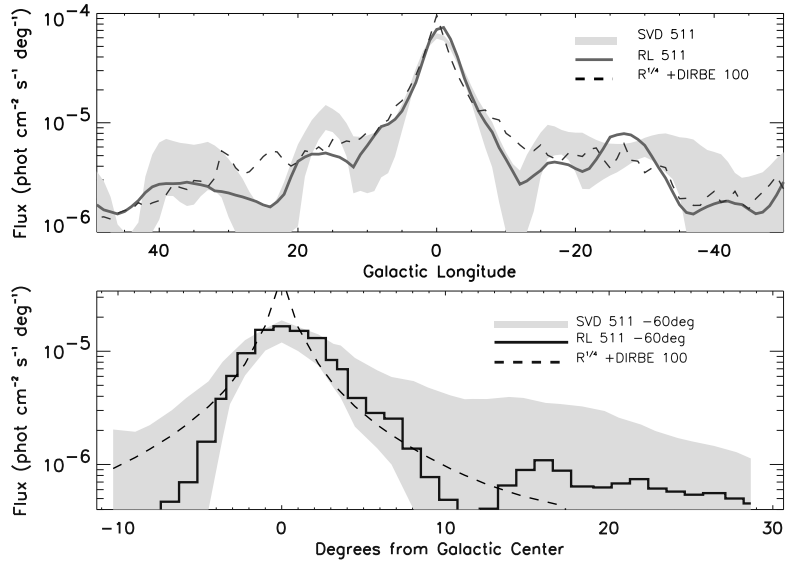, width=11cm}}
\vspace{2mm}
\caption{Figure 3a shows the integrated intensity from the
$|b|$ $\leq$ 2$^{\circ}$ portion of the galactic plane. The SVD-511
cut is shown with 1$\sigma$ uncertainties, the RL-511 map and
the R$^{1/4}$ bulge and the DIRBE 100 disk model are shown without
uncertainties.
Figure 3b shows a 1$^{\circ}$ wide cut through the
PLE region, determined to be best characterized as extending at a
60$^{\circ}$ angle relative to the galactic plane. All maps were 
generated from the OSSE/SMM data-set.}
\end{figure}
                 
The lower panel of Figure 3 shows a cut through the 
PLE at an angle of 60$^{\circ}$ relative to the 
negative-longitude galactic plane (an angle suggested by the SVD map). 
As seen in the longitudinal cut, the R$^{1/4}$ shape is more 
centrally peaked. 
In the anti-PLE direction, all three maps agree, all falling smoothly. 
Both the SVD-511 and the RL-511 are brighter in the +3$^{\circ}$ 
$\rightarrow$ +8$^{\circ}$ region than the corresponding negative 
region. Beyond +8$^{\circ}$, the sensitivity becomes poor, and 
 there is little significance to the 
differences between the SVD and RL maps. 
 The PCONT map 
(not shown) does not
show a corresponding enhancement, though interpretation of this 
result as being due to annihilation physics, or alternatively being due 
to fitting systematics is not justified in this preliminary 
analysis.\footnote{The 
``annihilation fountain" model (Dermer \& Skibo 1997) would have a 
low positronium fraction due to the high temperature, as direct 
annihilation with free electrons dominates over radiative recombination
above 10$^{6}$K (Bussard, Ramaty \& Drachman 1979). However, the absence of 
broad emission in the TGRS spectra provides a 
constraint to this scenario (Harris et al. 1998).} 

\section{Model-fitting the Galactic Center Region}

The maps of 511 keV line emission support the PURC97 
representation of the galactic emission as being due to 
bulge, disk and PLE components. To quantify the contributions by the
bulge and disk components, three bulge shapes have been combined with 
28 disk shapes and compared with the OSSE, then the OSSE/SMM, then the 
OSSE/SMM/TGRS 511 keV data-sets. 
The bulge shapes tested are; a GC point source, a Gaussian, and
the projection of a truncated R$^{1/4}$ function. 
 Shown in the left panel of
Figure 4 are the chi-squared values for
best-fitting bulges of each shape paired with
28 disk models and fit to the OSSE
data-set. 
The disks have been ordered by the equivalent latitude FWHM of a Gaussian
profile.\footnote{The map references are:
DIRBE maps (Hauser et al. 1998), the CO map (Dame et al. 1987),
the Hot Plasma (Koyama et al. 1989), the R$^{1/4}$, HF Light \&
Disk Light (Higdon \& Fowler 1987), the M31 maps
(Ciardullo et al. 1987). ``90x10'', etc. 
refer to the FWHM of Gaussian disks.}
In the fits, the FWHM of the Gaussian bulges have been permitted to vary 
(best-fit FWHM values range from
3.9$^{\circ}$ to 5.7$^{\circ}$). The R$^{1/4}$ radial function has been 
truncated to a constant value
inside of a radius (R$_{min}$).\footnote{The R$^{1/4}$ function is
$\rho$(R $\leq$ 0.24) = A$\cdot$R$^{-6/8}$$\cdot$\{exp(-B$\cdot$R$^{1/4}$) \},
$\rho$(R $\geq$ 0.24) = 5/4$\cdot$A$\cdot$R$^{-7/8}$ \{exp(-B$\cdot$R$^{1/4}$)\
-C$\cdot$R$^{-1/4}$], where R = radial distance from GC in kpc, and
A,B,C are constants.}
  R$_{min}$ has been constrained to be between 50 pc and 700 pc. 
The radial distribution has then been projected
onto the line of sight
assuming the GC to be 8 kpc distant. For all R$_{min}$, the R$^{1/4}$
has extended wings relative to the Gaussian shape; for small R$_{min}$,
the R$^{1/4}$ is also more centrally peaked. For every disk tested,
the extended bulges are strongly favored over the GC point source. The
R$^{1/4}$ bulge shows a slight preference for thin disks,
while the Gaussian bulge shows
a stronger preference for thicker disks. This is interpreted as a hint
of emission separated from both the galactic plane and the GC.
 For the R$^{1/4}$ bulge shape, 
the extended bulge wings account for this
emission. For the Gaussian, a thicker disk is required.

\begin{figure}
\label{modfit}
\centerline{\epsfig{file=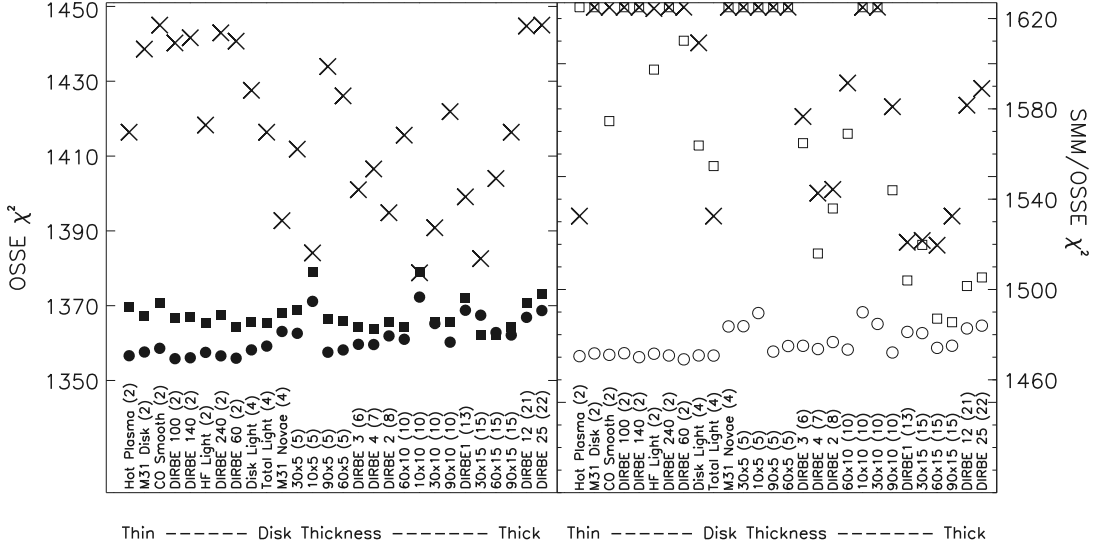, width=5.9in}}
\vspace{-.0cm}
\caption{Results of bulge-disk fitting to OSSE-only (left panel) and
OSSE/SMM (right panel) data-sets. The equivalent latitude 
FWHM of the disks are shown
in parentheses. ``90x10'', etc. refers to the FWHM of 
disks with a Gaussian profile 
in width and thickness. The R$^{1/4}$ bulges (filled and open
circles) and Gaussian bulges (filled and open squares) fit both
data-sets better than does the point source (X). When fitting the OSSE
data-set,
thin disks are favored for the R$^{1/4}$ shape, but disfavored
 for the Gaussian shape. The combined OSSE/SMM data-set
strengthens these tendencies.
Very large $\chi^{2}$ values for point source \& Gaussian fits
were truncated to 1622.}
\end{figure}

In the right panel of figure 4, the
SMM data is combined with the OSSE data. The fits for the R$^{1/4}$ solutions
are similar to the OSSE-only plots. However, only the thicker
disk-Gaussian bulge
solutions are able to approximate the quality of the R$^{1/4}$ fits.
As seen in Table 1, 
the B/D ratios are larger for R$^{1/4}$ solutions than for Gaussian
solutions. The total flux of the R$^{1/4}$ solutions change little
when SMM and then TGRS data is added, but for the Gaussian solutions the
total flux
can change considerably. For many Gaussian bulge-disk combinations, the OSSE
data would not permit insertion of the SMM or TGRS-required flux without
violating OSSE constraints, leading to poor fits. These solutions do not 
necessarily span the range of possible bulge-disk combinations, nor are they 
unique, but they do suggest ranges of plausible B/D ratios and total fluxes.
The uncertainties of the parameters are not shown in Table 1 due to 
concern about them being misinterpreted. The standard approach is to 
fix all but a single parameter, and calculate the degradation to the fit 
that results from varying that single parameter. The uncertainties that
result do not account for the effect of varying the other parameters, nor 
does it account for other potential model shapes. In this paper, the 
uncertainties are evident in Table 1, but are not calculated. 

\begin{table}
\caption{Fit parameters for
better-fitting models to the OSSE-only, OSSE/SMM \& OSSE/SMM/TGRS
data-sets (HP=Hot Plasma, DL=Disk Light).
Bulge emission is dominant for R$^{1/4}$ pairs, while disk
emission is dominant for Gaussian pairs.
The Richardson-Lucy and SVD map parameters are shown for comparison.}
\vspace{1mm}
\begin{tabular}{llccclccclccc}
Models &
\multicolumn{4}{c}{OSSE-only} &
\multicolumn{4}{c}{OSSE/SMM} &
\multicolumn{4}{c}{OSSE/SMM/TGRS}\\
\tableline
{\bf R$^{1/4}$} &R$_{min}$&B/D & F$_{T}$\tablenote{The total flux in
units of 10$^{-4}$
phot cm$^{-2}$ s$^{-1}$.} & $\chi^{2}/\alpha$ & R$_{min}$&
 B/D & F$_{T}$$^{a}$ & $\chi^{2}/\alpha$ &R$_{min}$& 
B/D & F$_{T}$$^{a}$ &
$\chi^{2}/\alpha$ \\
HP& 100pc & 1.6 & 25.7 & 1.18 & 200pc & 1.7 & 28.3 & 1.20 &
200pc & 1.2 & 27.6 & 1.20 \\
CO& 50pc & 2.6 & 30.0 & 1.18 & 50pc & 1.2 & 28.9 & 1.20 &
50pc & 3.3 & 31.3 & 1.20 \\
DL& 50pc & 2.7 & 25.2 & 1.18 & 50pc & 1.5 & 29.0 & 1.20 &
50pc & 0.8 & 28.2 & 1.20 \\
90x10 & 50pc & 0.6 & 27.4 & 1.18 & 50pc & 0.9 & 27.9 & 1.20 &
50pc & 0.6 & 26.8 & 1.20 \\
\tableline
{\bf Gaus.} &Fwhm& B/D & F$_{T}$$^{a}$ & $\chi^{2}/\alpha$ &Fwhm&
 B/D & F$_{T}$$^{a}$ & $\chi^{2}/\alpha$ &Fwhm& B/D & 
F$_{T}$$^{a}$ &
 $\chi^{2}/\alpha$ \\
60x15 & 4.4$^{\circ}$ & 0.2 & 20.9 & 1.19 &
4.3$^{\circ}$ & 0.2 & 26.4 & 1.21 &
4.3$^{\circ}$ & 0.2 & 26.4 & 1.21 \\
90x15 & 5.1$^{\circ}$ & 0.3 & 21.6 & 1.19 &
4.8$^{\circ}$ & 0.2 & 29.6 & 1.32 &
5.7$^{\circ}$ & 0.2 & 29.9 & 1.21 \\
DRB12 & 5.5$^{\circ}$ &0.1 & 29.9 & 1.28 &
5.7$^{\circ}$ & 0.2 & 31.3 & 1.33 &
5.7$^{\circ}$ & 0.2 & 31.4 & 1.22 \\
\tableline
\multicolumn{2}{l}{{\bf RL Map}}\tablenote{The degrees of freedom have been 
set equal to those of the models.} & & 38.3 & 1.13 & & 0.63 & 37.0 & 1.14 &
& & & \\
\multicolumn{2}{l}{{\bf SVD Map}}$^{b}$ 
& & 31.1 & 1.15 & & 0.49 & 25.8 & 1.18 &
& & & \\
\tableline
\end{tabular}
\end{table}

The results shown in Table 1 have ignored the 
existence of a PLE and its potential influence upon the B/D and 
F$_{Tot}$ parameters. PURC97 
quantified the PLE flux 
by two methods; (1) by subtracting the mirror region from 
the outputs of the mapping techniques, and (2) by fitting the data-set 
with three components (bulge, disk and PLE), all with Gaussian 
profiles. The two methods yield 
very different PLE flux 
results. The PURC97 maps suggest the PLE flux to be (1.3-1.5) x 10$^{-4}$ 
phot cm$^{-2}$ s$^{-1}$. The 3-Gaussian fitting suggests  
(5.4-8.8) x 10$^{-4}$ phot cm$^{-2}$ s$^{-1}$. The current maps 
suggest (1.2$\pm$0.5) x 10$^{-4}$ phot cm$^{-2}$ s$^{-1}$, in agreement
with the PURC97 maps. Quantifying the PLE by mirror-region 
subtraction, the B/D ratios 
for the RL and SVD maps are 0.63 and 0.49, as shown in Table 
1. The B/D values are intermediate to the R$^{1/4}$ and Gaussian solutions,
suggesting a general agreement between mapping and two component modeling.

The PURC97 3-Gaussian solution, when inserted with quoted parameters, 
is not an acceptable solution for the current OSSE/SMM/TGRS data-set 
($\chi^{2}$/$\alpha$ = 1.34), 
but the same method can be applied to examples of 
better-fitting models from Table 1. When a Gaussian representing 
the PLE
is added to a 4.3$^{\circ}$ Gaussian bulge + 60$^{\circ}$x15$^{\circ}$
Gaussian disk, and the B/D re-optimized,
the B/D rises from 0.163 to 0.169 ($\Delta\chi^{2}$=-16.4). The PLE centroid
for this Gaussian fit is determined to be ($l,b$) = -2$^{\circ}$, +8$^{\circ}$,
and the PLE flux is 1.1 x 10$^{-4}$ phot cm$^{-2}$
s$^{-1}$. When a PLE Gaussian is added to the R$^{1/4}$
+CO Smooth model, the B/D lowers slightly
from 3.3 to 2.6 ($\Delta\chi^{2}$=-12.3), and the parameters become:
($l,b$) = -2$^{\circ}$, +1$^{\circ}$, PLE flux = 0.7 x 10$^{-4}$
phot cm$^{-2}$ s$^{-1}$. These values are given not to suggest a
refinement of the PLE, but rather to demonstrate that the B/D depends
more on the bulge shape than on the PLE (and vise versa).
The current modeling
of the PLE lowers the flux values to better agreement with the mapping
fluxes, but the interpretation must be that the characteristics of
any possible
PLE are too poorly constrained and too model dependent to claim flux
values and emission centroids (and the corresponding uncertainties).

\section{Discussion}

OSSE investigations of galactic positron annihilation have recently 
been 
expanded into three new areas; utilization of archival data near the 
GC, utilization of archival data away from the GC and mapping the 
positronium continuum emission. 
This expansion makes the results more dependent upon the spectral 
analysis, which is currently in its preliminary stage. 
 The portion of this analysis least dependent upon these new 
complications are the 511 keV line flux values. Although the spatial 
coverage of maps from these values
 has increased, characterization of the emission with 
three components (bulge, disk and PLE) remains adequate. 
An extended bulge emission has been shown to 
be favored over point-like emission from the GC for a collection 
of 28 disk models. Comparing two approximations of the bulge shape, 
Gaussian bulges
require thick disks to fit the data-sets, while R$^{1/4}$ bulges are
less dependent upon the disk thickness, slightly favoring thin disks.
The bulge shape must be better resolved with future measurements
to improve the constraints upon the B/D ratio (currently ranging 
from 0.2 -3.3 for the combinations tested). 
The total flux is better constrained, ranging between
(20.9-31.4) x 10$^{-4}$ phot cm$^{-2}$ s$^{-1}$. The effect of a
potential PLE upon these parameters is shown to be less than the
bulge-shape uncertainty when the PLE is approximated by a Gaussian.

Positron astronomy is 30 years old but remains in its infancy. Of 
current generation gamma-ray telescopes, OSSE is the best-suited to 
investigate the sky distribution of this emission.  The time allotted 
to OSSE annihilation studies has increased in recent cycles of 
the CGRO mission. This time, combined with the expanded OSSE analysis, 
and the impending launch of the INTEGRAL telescopes should produce 
significant improvements of our understanding of the nature of galactic 
positron annihilation.

\end{document}